\documentstyle[12pt,epsf,a4]{article}
\begin{document}
\pagestyle{empty}
\parsep  6pt plus 1pt minus 1pt
\parindent 12truemm
\def\beq{\begin{equation}}
\def\eeq{\end{equation}}
\newcommand{\bea}{\begin{eqnarray}}
\newcommand{\eea}{\end{eqnarray}}
\newcommand{\nn}{\nonumber}
\def\simleq{\; \raise0.3ex\hbox{$<$\kern-0.75em
      \raise-1.1ex\hbox{$\sim$}}\; }
\def\simgeq{\; \raise0.3ex\hbox{$>$\kern-0.75em
      \raise-1.1ex\hbox{$\sim$}}\; }
\def\noi{\noindent}
\font\boldgreek=cmmib10
\textfont9=\boldgreek
\mathchardef\myrho="091A
\def\bfrho{{\fam=9 \myrho}\fam=1}
\def\R{ {\rm R \kern -.31cm I \kern .15cm}}
\def\C{ {\rm C \kern -.15cm \vrule width.5pt
\kern .12cm}}
\def\Z{ {\rm Z \kern -.27cm \angle \kern .02cm}}
\def\N{ {\rm N \kern -.26cm \vrule width.4pt \kern .10cm}}
\def\1{{\rm 1\mskip-4.5mu l} }

\vbox to 2 truecm {}
\centerline{\large \bf Duality in the non-relativistic harmonic oscillator quark model} \par \vskip 3 truemm  
\centerline{\large \bf in the Shifman-Voloshin limit : a pedagogical example}\par \vskip 2 truecm 

\centerline{A. Le Yaouanc$^{a}$, D. Melikhov$^{b}$ \footnote{Alexander-von-Humboldt fellow. On leave from Nuclear Physics Institute, Moscow State University, Moscow, Russia}, V. Morenas$^{c}$, L. Oliver$^{a}$, O. P\`ene$^{a}$ and J.-C. Raynal$^{a}$} \par \vskip 3 truemm

$^a$ Laboratoire de Physique Th\'eorique, Universit\'e de Paris XI, \par 
B\^atiment 210, 91405 Orsay Cedex, France\footnote{
Unit\'e Mixte de Recherche - CNRS - UMR 8627}\\

$^b$ Institut f\"ur Theoretische Physik, Universit\"at Heidelberg, \par
Philosophenweg 16, D-69120, Heidelberg, Germany\\

$^c$ Laboratoire de Physique Corpusculaire, Universit\'e Blaise Pascal - CNRS/IN2P3,\par 63000
Aubi\`ere Cedex, France
\par \vskip 1 truecm

\begin{abstract}
The detailed way in which duality between sum of exclusive states and the free quark model
description operates in semileptonic total decay  widths, is analysed. It is made very
explicit by the use of the non relativistic harmonic oscillator quark model in the SV
limit, and a simple interaction current with the lepton pair. In particular, the Voloshin
sum rule is found to eliminate the  mismatches of order $\delta m/m_b^2$. \end{abstract}

\par \vskip 3 truecm
\noindent LPT-Orsay 00-29 \par 
\noindent HD-THEP-00-14 \par
\noindent PCCF RI 00-10 \par
\noindent hep-ph/0005039

\newpage
\pagestyle{plain}
\section{Introduction}
\hskip \parindent Discussions have recently arisen about the possibility that
expectations from OPE for some types of semi-leptonic rates may be violated by
terms of order $1/m_Q$. The argument of Nathan Isgur \cite{isgur} is founded
on general considerations ; namely the duality is obtained in 
the infinite mass limit through cancellation between 
the falloff of the ground state contribution and the rise of
the excitations (the Bjorken sum rule indeed relates the derivatives
of these contributions with respect to $w$, near $w=1$), but at finite mass
there is some mismatch near zero recoil, which could be of
order $1/m_Q$. Indeed, in terms of $t$, the
quadri-dimensional transfer\footnote{$t$ is  we use the old standard notation $t$, to avoid confusion with
the tridimensional $|\vec{q}{\,}|^2$, which will be used extensively in this non
relativistic (NR) context.} :
\bea
t=(q^0)^2-\vec{q}~^2 , 
\eea
the respective $t_{max}$ do not coincide anymore. The argument
is then given by the author further likeliness by some calculations within a very simple
``toy'' model :  the non relativistic  harmonic oscillator (HO) potential model. \par
In the present letter, we will not discuss directly the issue
about QCD (see our article \cite{QCD}). We simply stick to the very model used in \cite{isgur}, and
show that within this model, calculating the {\it total integrated} rate $\Gamma_{inclusive}$ by summation on the relevant final (bound) states, duality with free quark decay rate is in fact 
satisfied, in
the SV (Shifman-Voloshin\cite{SV}) limit\footnote{By SV limit, we do not mean simply 
that the recoil velocity is small, but also, as in the original paper \cite{SV}, that ${\delta m
\over m}$ is small ; in addition, $\delta m$ is taken large with respect to light quark
parameters ; in non relativistic quantum mechanics, we assume : $\Delta={1 \over
m_d R^2}\ll \delta m~and~m_d, \delta m \ll m_b,m_c$, $\Delta$ being the level spacing.}; this means that the difference $\Gamma_{inclusive}-\Gamma_{free~quark}$ comes out as expected, which implies in particular (as discussed
below) cancellation of terms of relative order $(\delta m)^2/m_b^2$ and $\delta
m/m_b^2$ (by relative, we mean with respect to the free quark decay rate ; note that such terms correspond to $(1/m_Q)^0,(1/m_Q)^1 $ in the usual $1/m_Q$ expansion). Our argument is for integrated decay rates, so we do not claim anything
on possible effects in differential or partially integrated rates. Also, of course, we
cannot exclude by such argument that something odd may happen in QCD.\par

One very interesting point raised in the discussion of
\cite{isgur} is about the very specific cancellations which are
necessary for duality to hold, and about the contributions of the various
regions of phase space.  
We try to analyze through our
demonstration how such cancellations occur in subleading order for total widths.
An interesting consequence of the analysis is 
that to find the required cancellations, one needs not to consider only the sum rule of Bjorken ; one has to take into account in addition the Voloshin sum rule
(the fact that one needs the sum rules has been
suggested by the Minnesota group in their discussion with N. Isgur \cite{isgur}, but is made here quite explicit ; for related discussions in QCD by the same group, see \cite{minnesota}). 
In fact, the Voloshin sum rule is exactly what is needed for cancellation of terms of relative order $\delta m/m_b^2$
in the difference $\Gamma_{inclusive}-\Gamma_{free~quark}$. The sum rules are trivially satisfied in the HO model, but it is not so trivial in general.
Our conclusion is not in contradiction with the mismatch occuring near zero
recoil, considered in \cite{isgur}, because the latter is very small parametrically
with respect to the terms we consider in the total width. 
\par

Note that the use of SV limit is not
essential to demonstrate duality in this way, and neither is the use 
of an HO potential. Their choice is pedagogical. Indeed we have also done the demonstration for
an arbitrary potential (\cite{articleNR})
and also for fixed $m_c/m_b$ ratio. Nevertheless, the particular case considered here is
of pedagogical interest, because on the one hand the discussion in the SV limit is much
simpler, and the similar discussion in QCD can  hardly be made beyond the SV limit, and
because on the other hand, within HO model, we can give explicit expressions. Moreover, we
are able to give a complete proof that in the HO model $1/m_Q$ terms are absent in the
ratio $\Gamma_{inclusive}/\Gamma_{free~quark}$ beyond the SV limit (article to appear \cite{JC}).
Note also that the demonstration is independent of the leptonic tensor, as we have also
shown elsewhere, but we choose here one specific for illustration. On the other hand, the
coefficient of the terms of order ${1 \over R^2 m_b^2}$, which we also evaluate, is
model-dependent (in particular it depends on the choice of the leptonic tensor; we choose
here one for illustration).

\section{Model}
\hskip \parindent $\bullet$ The model for hadrons is the non relativistic harmonic oscillator quark model
(the motion of quarks both internal and due to
overall hadron are both treated non relativistically),
describing the initial (quarks $b$ and $d$) and final ($c$ and $d$) hadrons.
The potential is assumed to be {\it flavor independent}, which is crucial for the demonstration. The great advantage of the harmonic oscillator, which appears in the summation on final states, is that very few states contributes to the transition rates in the limited expansion in ${1 over m_b}$ which we perform (see next section).
Energy levels, for a state labelled by $(n_x,n_y,n_z)$, $n=n_x+n_y+n_z$, write :
\bea
E_n=m_{b,c}+m_d+\left({3\over 2}+n \right) {1 \over \mu_{b,c} R_{b,c}^2}
\eea
where $\mu_{b,c}$ are the reduced masses ${m_{b,c} m_d\over m_{b,c}+m_d}$ and the radii $R_{b,c}^2$
can be written as :
\bea
R_{b,c}^2=\sqrt{m_d \over \mu_{b,c}}~R_{\infty}^2
\eea
$R_{\infty}$ being the radius in the infinite mass limit. We will often denote the first level excitation energy  
in the infinite mass limit as :
\bea
\Delta={1 \over m_d~R_{\infty}^2} \label{Delta} .
\eea
For simplicity, from now  on, we denote : 
\bea 
R_{\infty}=R 
\eea
\par
$\bullet$ Quarks are then coupled to lepton pairs : $b \to c \ell \nu$, 
through a quark vector current $j^0=1$, $\vec{j}=0$ \footnote{Note that we do not claim to make a systematic non relativistic {\it expansion} of a relativistic theory, but only to consider a non relativistic Hamiltonian for the bound states; we can choose freely the weak interaction current. The essential point is then to treat consistently the matrix elements according to the chosen interactions, in the specified SV expansion.}
(or equivalently we can speak of spinless quarks), and a leptonic tensor, which will be
described by functions denoted generically through letter $L$ and some arguments and
indices. $\vec{P}$ and $\vec{P'}$ are the initial and final hadron momenta ; the total
energies of hadrons are $P^0=E+\vec{P}{\,}^2/2(m_b+m_d),
P'^0=E'+\vec{P'}{\,}^2/2(m_c+m_d)$, with $E,E'$ the energies at rest ; but, in practice, we will always work in the initial
hadron rest frame : $\vec{P}=0$ ;  $\vec{P}$ and $\vec{P'}$ are the initial and final
quark momenta. We denote  $\vec{q}=\vec{P}-\vec{P'}=-\vec{P'}$. The basic equation is
then, in the initial state rest frame :

\bea  \Gamma_{inclusive}=K \sum_n
\int_0^{|\vec{q}{\,}|_{max,n}}~d|\vec{q}{\,}|~|\vec{q}{\,}|^2~L_n(|\vec{q}{\,}|)~\sum_{n=n_x+n_y+n_z}|j_{0
\to (n_x,n_y,n_z)}|^2. \eea 

\noindent The constant $K$ depends  only on the decay interaction
strength. The constant $K$  will be  ommitted in the rest of the letter.
$\sum_{n=n_x+n_y+n_z}|j_{0 \to (n_x,n_y,n_z)}|^2$ only depends on $|\vec{q}{\,}|$. The
angular integration has been performed. The notations $L_n(|\vec{q}{\,}|)$ and $|\vec{q}{\,}|_{max,n}$ are
explained now. A priori, after angular integration, the leptonic tensor appears through a
function of energy loss $q^0$ and $\vec{q}{\,}^2$, $L(q^0,|\vec{q}{\,}|^2)$. However, for the
decay from the ground state to a h.o. state labelled by $(n_x,n_y,n_z)$,  by energy
conservation, $q^0=P^0-P'^{0}$ is just a function of $|\vec{q}{\,}|$ and $(n_x,n_y,n_z)$. Moreover,
the energy loss $q^0$ will  depend only on $n=n_x+n_y+n_z$. and we then denote as $L_n(|\vec{q}{\,}|)$
the result of $L(q^0,|\vec{q}{\,}|^2)$, when the energy loss $q^0$ is assumed to be calculated for the
corresponding $n$, as a function of $|\vec{q}{\,}|$. Indeed, for a state with degree of
excitation $n$ : 

\bea q^0(n,|\vec{q}{\,}|)=m_b-m_c+{3\over 2}~({1 \over \mu_b R_b^2}-{1
\over \mu_c R_c^2}) -n{1 \over \mu_c R_c^2}-{|\vec{q}{\,}|^2 \over 2 (m_c+m_d)}. \eea

\noindent Now $q_{max}$ is determined by the equation $t=(q^0)^2-|\vec{q}{\,}|^2=0$,
$q^0(|\vec{q}{\,}|)=|\vec{q}{\,}|$ : 

\bea |\vec{q}{\,}|_{max,n}={2 (m_c+m_d) (\delta E)_n
\over 2 (m_c+m_d)+\sqrt{(m_c+m_d)^2+2 (m_c+m_d) (\delta E)_n}} \eea

\noindent  where  

\bea (\delta
E)_n=q^0(n,\vec{q}=0~)=m_b-m_c+{3\over 2}~({1 \over \mu_b R_b^2}-{1 \over \mu_c R_c^2})
-n{1 \over \mu_c R_c^2}. \eea 

\noindent $|\vec{q}{\,}|_{max}$ just depends on $n$.
$L(q^0,|\vec{q}{\,}|^2)$ can be taken as an arbitrary function without spoiling any of the
general statements made below, but for definiteness we will henceforth choose~:  \bea 
L(q^0,\vec{q}~^2)=3 (q^0)^2-|\vec{q}{\,}|^2 ,
\eea
inspired by a static quark approximation of the V-A current.
The corresponding free quark decay rate is :
\bea
\Gamma_{free}=K~\int_0^{|\vec{q}{\,}|_{max,free}}~d|\vec{q}{\,}|~|\vec{q}{\,}|^2~L(q^0,|\vec{q}{\,}|^2)
\eea
with :
\bea
q^0(free,|\vec{q}{\,}|)=m_b-m_c-{|\vec{q}{\,}|^2 \over 2 m_c}
\eea
\bea
|\vec{q}{\,}|_{max,free}={2 m_c \delta m \over (m_c+\sqrt{m_c^2+2 m_c \delta m })}
\eea
with $\delta m =m_b-m_c$.\par

\section{SV expansion and demonstration of duality}
\hskip \parindent $\bullet$ We have then to consider the expansion of
\bea
\epsilon={\Gamma_{inclusive}-\Gamma_{free} \over \Gamma_{free}}
\eea
in powers of ${1 \over m_b}$, and the aim is in principle to show that it begins 
with order ${1 \over m_b^2}$ only, as expected from a formal OPE (the NR version of OPE will be explained in the more developped article). More precisely this holds in the limit $m_b
\to \infty$ with $r={m_c \over m_b}$ fixed, for which we reserve for clarity the term " usual $1/m_Q$ expansion". However, we will work in the SV (Shifman-Voloshin) limit, which corresponds to making in addition an expansion in $1-r$. Namely, with :
\bea
\delta m=m_b-m_c , 
\eea
we write $m_c=m_b-\delta m$  and we expand in powers of ${1 \over m_b}$, keeping
$\delta m$ fixed, as well as the light quark parameters, $m_d$, $1/R$ ; then,
we make a second limited expansion, taking $\Delta={1 \over m_d R^2}$ small with respect to $\delta m$. The terms have the form ${(\delta
m)^{k'} \over (m_b)^k}$ times light quark factors. But then the aim must be more than just showing the absence of powers \footnote{Note that, in the present letter, we term generically
as {\it power} ${1 \over (m_b)^k}$ all the terms which contain the factor ${1 \over (m_b)^k}$,
whatever the powers of $\delta m$ and light quantities.} ${1 \over (m_b)^k}$, $k < 2$ in $\epsilon$.\par

Indeed, if it is true, this would not in principle preclude terms of the type ${(\delta
m)^{k'} \over m_b^2}$ ($k' > 0$) in $\epsilon$. Such terms would be large in practice, since
$\delta m$ is not so small. And in fact, they would correspond, in terms of the usual $1/m_Q$ expansion, to terms of order $(1/m_Q)^0,(1/m_Q)$, since $\delta m$ would be then $\propto m_Q$. Such terms are not
expected from OPE. We must therefore show that such terms do not exist in the final result, and we
show it in fact. More precisely, we show  that potentially large terms of the type
${(\delta m)^2 \over m_b^2}$, ${m_d \delta m \over m_b^2}$, which appear in particular
contributions, do finally cancel out, leaving us with
terms of the type ${1 \over R^2 m_b^2}$ (terms with $k' > 2$ simply do not appear in the
way  we calculate $\epsilon$, neither do terms with power ${1 \over (m_b)^0}$ or ${1 \over m_b}$ - in fact, the delicate part consists in showing the cancellation of ${m_d \delta m \over m_b^2}$ terms). This is all that is required by duality with free quarks, as concerns the terms with power $({1 \over m_b})^k$, $k \leq 2$. We will calculate
the  terms of type ${1 \over R^2 m_b^2}$, which do not vanish in general. Note that such terms are small with respect to ${m_d \delta m \over m_b^2}$ by a factor ${\Delta \over
\delta m}$. In the usual $1/m_Q$ expansion they correspond to order $1/m_Q^2$. On the other hand, we will not calculate in the
expansion of $\epsilon$ smaller terms having also the power ${1 \over m_b^2}$, but which contain still additional powers of ${\Delta \over
\delta m}$ with respect to ${1 \over R^2 m_b^2}$, corresponding in
$\Gamma_{inclusive}-\Gamma_{free}$ to terms like ${(\delta m)^4 \times\Delta \over m_b^2}$,
${(\delta m)^3\times\Delta^2 \over m_b^2}$, etc... and retain only the terms proportional
to $\Gamma_{free} \propto (\delta m)^5$. The neglected terms correspond to terms of relative order $1/m_Q^3$ or beyond in the $1/m_Q$ expansion. For sake of simplicity, we will neither examine further checks of duality in terms of the type ${(\delta m)^{k'} \over (m_b)^k}$, with $k > 2$. \par 
In any case, we see that we do have to
calculate terms with a power ${1 \over m_b^2}$ and not ${1 \over m_b}$ only, since the terms with a power ${1 \over m_b^2}$ may correspond to terms of the order $(1/m_Q)^0,(1/m_Q)^1$ in the usual expansion. The method
precisely consists in writing the difference $\Gamma_{inclusive}-\Gamma_{free}$ as a sum
of terms which contain already a power ${1 \over m_b^2}$, and then to demonstrate the above 
additional cancellations.\par

	$\bullet$ The advantage of harmonic oscillator (HO) model is that the level $n=1$ (which
corresponds to $L=1$ states) appears only with a power ${1 \over m_b^2}$, and that higher
levels come only with a power ${1 \over m_b^3}$ at least. Since we keep terms with a power
${1 \over m_b^i}$,$i \leq 2$ , we only need consider $n=0,1$ states. For sake of simplicity, we denote their respective contributions $\Gamma_{0,1}$.\par
We have at this order, by expanding the matrix elements :
\bea
\Gamma_0 \simeq \int_0^{|\vec{q}{\,}|_{max,0}}~d|\vec{q}{\,}|~|\vec{q}{\,}|^2~L_{n=0}(|\vec{q}{\,}|)~(1-\rho^2~{|\vec{q}{\,}|^2 \over m_b^2}),
\eea
where $\rho^2={1 \over 2}m_d^2 R^2$ is the standard slope of the 
ground state form factor with respect to $w$  
($w-1 \simeq {1 \over 2}{|\vec{q}{\,}|^2 \over m_b^2}$) ; note that effect of non complete overlapping between hadrons with $b$ and $c$ quarks is completely negligible here, since it contributes at order $1/R^2 {(\delta m)^2 \over m_b^4}$. For $L=1$ states :
\bea
\Gamma_1 \simeq \int_0^{|\vec{q}{\,}|_{max,1}}~d|\vec{q}{\,}|~|\vec{q}{\,}|^2~L_{n=1}(|\vec{q}{\,}|)~\tau^2~{|\vec{q}{\,}|^2 \over m_b^2},
\eea
with $\tau={m_d R \over \sqrt{2}}$ corresponding to the $\tau_{1/2,3/2}(w=1) \times \sqrt{3}$. The other excitations do not contribute at this order, because the matrix element $<n|\vec{r}|0>$ is non zero only if $n=1$.
From the explicit expressions, we have the relations :
\bea
\rho^2-\tau^2=0,\label{bjorken}
\eea
\bea
\Delta~\tau^2={m_d \over 2}, \label{voloshin}
\eea
($\Delta$ being the level spacing, \ref{Delta}) as non relativistic analogues of the Bjorken and Voloshin sum rules. These
sum rules could then be used to generalise the present analysis.\\
In fact, we will try as much as possible not to specify separately
$\Delta,\rho,\tau$, but to use only the above sum rules and expressions
for $\Gamma_{0,1}$.\par
	$\bullet$ The strategy is to note that the difference between 
$\Gamma_0 +\Gamma_1$ and $\Gamma_{free}$ can be reexpressed by successive steps~:\par
1) Decomposition into the same difference with $L_{0,1}(|\vec{q}{\,}|)$ substituted by their free counterpart
$L_{free}(|\vec{q}{\,}|)$ (contribution (I)) plus a ${1 \over m_b^2}$ term (contribution (II)). 
2) Then the first contribution (I) is 
rewritten trivially as a difference between two contributions having a power ${1 \over m_b^2}$ relative to the free quark decay integrand, further shown to be of relative order ${1 \over R^2 m_b^2}$ or smaller.
3) It is also shown that in contribution (II), which contains manifestly a power ${1 \over m_b^2}$, there are  only terms of the type ${1 \over R^2 m_b^2}$ or smaller.\par
	
	$\bullet$ In a first step, using the respective expressions
given above for the $q^0(|\vec{q}{\,}|)$'s, and expanding it to the required order, we find :
\bea
q^0(free,|\vec{q}{\,}|) \simeq  \delta m -{|\vec{q}{\,}|^2 \over 2 m_c}
\eea
\bea
q^0(n=0,|\vec{q}{\,}|) \simeq \delta m (1-{3 \over 4 m_b^2 R^2})-{|\vec{q}{\,}|^2 \over 2(m_c+m_d)}
\eea
\bea
q^0(n=1, |\vec{q}{\,}|)\simeq \delta m -\Delta
\eea
Note that in our expansion, the main term in the three quantities is $\delta m$. The main term of $q_{max}$ is then also $\delta m$ ($q_{max}=q^0$ at $t=0$).
\noindent We use these expansions to make, in the integrals for $\Gamma_{0,1}$ : 
\bea
L_0(|\vec{q}{\,}|) \simeq L_{free}(|\vec{q}{\,}|)+
6 \delta m (-{3 \delta m \over 4 R^2 m_b^2}+{m_d~|\vec{q}{\,}|^2 \over 2 m_b^2})
\eea
\bea
L_1(|\vec{q}{\,}|) \simeq L_{free}(|\vec{q}{\,}|)+6\delta m(-\Delta)+3~\Delta^2
\eea
The second terms in the r.h.s. come from the difference between the respective
$q^0$, as a function of $|\vec{q}{\,}|$.
Note that in the expansion of $L_1(|\vec{q}{\,}|)$, one can neglect terms in
${1 \over m_b^2}$ because $\Gamma_1$ has already a power ${1 \over m_b^2}$.
We get $\Gamma_{inclusive}-\Gamma_{free} \simeq \delta \Gamma_{I}+\delta \Gamma_{II}$ with :
\bea
\delta \Gamma_{I}=&\int_0^{|\vec{q}{\,}|_{max,0}}&~d|\vec{q}{\,}|~|\vec{q}{\,}|^2~L_{free}(|\vec{q}{\,}|)~(1-\rho^2~{|\vec{q}{\,}|^2 \over m_b^2})+\cr
&\int_0^{|\vec{q}{\,}|_{max,1}}&~d|\vec{q}{\,}|~|\vec{q}{\,}|^2~L_{free}(|\vec{q}{\,}|)~\tau^2~{|\vec{q}{\,}|^2 \over m_b^2}-\cr
&\int_0^{|\vec{q}{\,}|_{max,free}}&~d|\vec{q}{\,}|~|\vec{q}{\,}|^2~L_{free}(|\vec{q}{\,}|)\,
\eea
and
\bea
\delta \Gamma_{II}=& \int_0^{|\vec{q}{\,}|_{max,0}}&~d|\vec{q}{\,}|~|\vec{q}{\,}|^2~
6\delta m (-{3 \delta m \over 4 R^2 m_b^2}+{m_d |\vec{q}{\,}|^2 \over 2 m_b^2})+\cr
&\int_0^{|\vec{q}{\,}|_{max,1}}&~d|\vec{q}{\,}|~|\vec{q}{\,}|^2~
(6\delta m(-\Delta)+3~\Delta^2)(\tau^2~{|\vec{q}{\,}|^2 \over m_b^2}).
\eea
\par
	
	$\bullet$ Contribution I. One can write it as
the difference of two integrals which {\it have already manifestly a factor} ${1 \over  m_b^2}$, i.e. the terms with power ${1 \over  (m_b)^0}$ or ${1 \over  m_b}$ are already cancelled (this amounts to using $\rho^2-\tau^2=0$, which is the particular form of the Bjorken sum rule in the model):
\bea
\delta \Gamma_{I}=&\int_{|\vec{q}{\,}|_{max,free}}^{|\vec{q}{\,}|_{max,0}}&~d|\vec{q}{\,}|~|\vec{q}{\,}|^2~L_{free}(|\vec{q}{\,}|) \ -\cr
&\int_{q_{max,1}}^{|\vec{q}{\,}|_{max,0}}&~d|\vec{q}{\,}|~|\vec{q}{\,}|^2~L_{free}(|\vec{q}{\,}|)~\tau^2~{|\vec{q}{\,}|^2 \over m_b^2}
\eea
We first expand each integral.
- One has :
\bea
|\vec{q}{\,}|_{max,0}-|\vec{q}{\,}|_{max,free} \simeq \delta m~({1 \over 2}{m_d \delta m \over m_b^2}-
{3 \over 4}{1 \over R^2 m_b^2}),
\eea
whence 
\bea
\int_{|\vec{q}{\,}|_{max,free}}^{|\vec{q}{\,}|_{max,0}}~d|\vec{q}{\,}|~|\vec{q}{\,}|^2~L_{free}(|\vec{q}{\,}|) \simeq 
\delta m~({1 \over 2}{m_d \delta m \over m_b^2}-{3 \over 4}{1 \over R^2 m_b^2})
(\delta m)^2~L_{free}(|\vec{q}{\,}|_{max}= \delta m) \label{first}. 
\eea
One can make $|\vec{q}{\,}| \simeq \delta m$ in the integrand, because the integration interval contains already a power $1/m_b^2$, and the difference between $|\vec{q}{\,}|$ and  $\delta m$ contains a further $1/m_b$ factor. \par
- The second integral is more delicate, because the integration interval has  not a factor $1/m_b^2$; it is just $\simeq \Delta$ ; the variation of $|\vec{q}{\,}|$ is not negligible. One must do a limited expansion of the integrand in powers of $\Delta \over \delta$, so as to retain at least terms of the type ${1 \over R^2 m_b^2}$. It is there that the second expansion,in powers of $\Delta \over \delta$, enters the game  :
\bea
\int_{|\vec{q}{\,}|_{max,1}}^{|\vec{q}{\,}|_{max,0}}~&&d|\vec{q}{\,}|\ |\vec{q}{\,}|^2~L_{free}(|\vec{q}{\,}|)~\tau^2~{|\vec{q}{\,}|^2 \over m_b^2} \simeq \cr 
\int_{\delta m - \Delta}^{\delta m}~&&d|\vec{q}{\,}| \ |\vec{q}{\,}|^2~L_{free}(|\vec{q}{\,}|)~\tau^2~{|\vec{q}{\,}|^2 \over m_b^2}\simeq \nonumber \\
&&\Delta (\delta m)^2~L_{free}(|\vec{q}{\,}| = \delta m) ~\tau^2~{(\delta m)^2 \over m_b^2}-\nonumber \\
&&{\Delta^2 \over 2}{\tau^2 \over m_b^2}~{d \over d|\vec{q}{\,}|} (|\vec{q}{\,}|^4~L_{free}(|\vec{q}{\,}|))(|\vec{q}{\,}|= \delta m). 
\label{second}
\eea
Let us note that to estimate the relative order of the different terms, one has to divide
by a reference rate, which will be taken to be the free quark decay rate ; now $(\delta
m)^3~L_{free}(|\vec{q}{\,}|= \delta m)$, as well as ${d \over d|\vec{q}{\,}|} (|\vec{q}{\,}|^4~L_{free}(|\vec{q}{\,}|))(|\vec{q}{\,}|= \delta m)$, are
of the order of the free quark decay rate (with our choice $L_{free}(|\vec{q}{\,}|= \delta m)\propto
(\delta m)^2$).\par  

Then, one can first observe that in fact not only all the terms
written in eq. (\ref{first}), (\ref{second}) have a relative power $1/m_Q^2$, but that they
are more precisely of relative order $m_d \delta m/m_b^2$ at most ; terms of relative
order $(\delta m)^2/m_b^2$ are already  cancelled. This will be obtained more generally
thanks to Bjorken sum rule.  Now, the term of relative order $m_d \delta m/m_b^2$
encountered in the r.h.s. of the first integral (\ref{first}) is cancelled  by the first
term in the r.h.s. of the second integral (\ref{second}), just using Voloshin sum rule (\ref{voloshin}), i.e. $\Delta \tau^2=m_d/2$. All the remaining contributions are of the type $(\delta m)^5~{1 \over R^2 m_b^2}$. We
can evaluate them readily and find them to cancel too {\it for the particular choice} made
for $L(q^0,|\vec{q}{\,}|)$. Finally : \bea \delta
\Gamma_{I}=&\int_{|\vec{q}{\,}|_{max,free}}^{|\vec{q}{\,}|_{max,0}}&~d|\vec{q}{\,}|~|\vec{q}{\,}|^2~L_{free}(|\vec{q}{\,}|) \ -\cr
&\int_{|\vec{q}{\,}|_{max,1}}^{|\vec{q}{\,}|_{max,0}}&~d|\vec{q}{\,}|~|\vec{q}{\,}|^2~L_{free}(|\vec{q}{\,}|)~\rho^2~{|\vec{q}{\,}|^2 \over m_b^2}
\simeq 0 \eea It must be emphasized that the cancellation can occur because the difference
between $|\vec{q}{\,}|_{max,n}$ and $|\vec{q}{\,}|_{max,free}$ is changing sign between the ground state and the
excitations. With our assumption $\Delta \ll \delta m$, one has $|\vec{q}{\,}|_{max,1}<|\vec{q}{\,}|_{max,free}<|\vec{q}{\,}|_{max,0}$\\

$\bullet$ Contribution II. It is also obvious that it contains already a power ${1 \over m_b^2}$.
On factorising $(\delta m)^5$, one sees that ${m_d \delta m \over m_b^2}$ terms are
present in the first integral (second term of the bracket in the integrand) :
$\int_0^{|\vec{q}{\,}|_{max,0}}~dq~q^2~ (6 \delta m {m_d q^2 \over 2 m_b^2}$)) and in the
second one (first term of the bracket in the integrand) :
$\int_0^{|\vec{q}{\,}|_{max,0}}~dq~q^2~(6 \delta m(-\Delta)(\tau^2~{|\vec{q}{\,}|^2 \over m_b^2})$)),
the rest  being smaller. It is easily seen that these ${m_d \delta m \over m_b^2}$
terms cancel at this order, just using Voloshin sum rule $\Delta \tau^2=m_d/2$, to leave a
smaller contribution, which is only of order ${1 \over R^2 m_b^2}$ ; the latter is found
by performing a limited expansion of the integrand as above eq. (\ref{first}) (the interval
is once more ${\cal{O}} (\Delta)$), in powers of $\Delta \over \delta m$  : 
\bea 3 {m_d \delta m \over
m_b^2}~\int_{|\vec{q}{\,}|_{max,1}}^{|\vec{q}{\,}|_{max,0}}~d|\vec{q}{\,}|~|\vec{q}{\,}|^4 \simeq 3~(\delta m)^5~{1 \over R^2 m_b^2}
\eea The other terms in the integrals are already manifestly of this order, and one ends
with : \bea \delta \Gamma_{II}={9 \over 5}~(\delta m)^5~{1 \over R^2 m_b^2}. \eea This
result has been checked by a systematic expansion using Mathematica.\par

Finally, with  $\Gamma_{free} \simeq {4 \over 5}~(\delta m)^5$:
\bea
\epsilon = {\Gamma_{0}+\Gamma_{1}-\Gamma_{free} \over \Gamma_{free}} 
\simeq {{9 \over 5}(\delta m)^5~{1 \over R^2 m_b^2} \over {4 \over 5}~(\delta m)^5} 
= {9 \over 4}{1 \over R^2 m_b^2}
\eea
Let us reinsist that it is of the order expected from OPE, unlike terms of
the type ${(\delta m)^2 \over m_b^2}$ or ${m_d \delta m \over m_b^2}$, which duely cancel, as has been shown.\par

\section{Relative magnitude of Isgur contribution}
\hskip \parindent $\bullet$ Let us now return briefly to the very discussion raised by
ref. \cite{isgur}. One could be worried why it is found there some duality violating effect, while we do not. The contradiction is only apparent. The answer seems to be that in {\it totally
integrated widths}, the effect  considered in \cite{isgur} is finally relatively small parametrically with
respect to the ones we have considered.  Let us show that. The mismatch
near zero recoil considered in \cite{isgur} is the integral of the ground state contribution over
$w_0(t)={m_B^2+m_D^2-t \over 2 m_B m_D}$ between $w_0(t=(m_B-m_{D})^2)=1$ and the
threshold for the excited state production $w_0(t=(m_B-m_{D^{**}})^2)$ ( the variable $w$
for the ground state contribution is considered as a function of $t$, $w_0(t)$). Let us
pass through the variable $\vec{q}$, which is more adapted to the NR problem, and denote
as $|\vec{q}{\,}|_n(t)$ the value of $|\vec{q}{\,}|$ which corresponds to some $t$ for a state $n$; the total
ground state contribution can be decomposed into two parts :  
\bea 
&&\Gamma_0 \simeq
\int_{|\vec{q}{\,}|_0(t=(m_B-m_D)^2)}^{|\vec{q}{\,}|_0(t=0)}~d|\vec{q}{\,}|~|\vec{q}{\,}|^2~L_{n=0}(|\vec{q}{\,}|)~(1-\rho^2~{|\vec{q}{\,}|^2 \over
m_b^2})= \nonumber \\
&&\int_{|\vec{q}{\,}|_0(t=(m_B-m_D)^2)}^{|\vec{q}{\,}|_0(t=0)}~d|\vec{q}{\,}|~|\vec{q}{\,}|^2~L_{n=0}(|\vec{q}{\,}|)-\int_{|\vec{q}{\,}|_0(t=(m_B-m_D)^2)}^{|\vec{q}{\,}|_0(t=(0)}~d|\vec{q}{\,}|~|\vec{q}{\,}|^2~L_{n=0}(|\vec{q}|{\,})~(\rho^2~{|\vec{q}{\,}|^2
\over m_b^2})    \nonumber \\.
\label{isgur00}
\eea 
In the infinite mass limit, $m_B-m_{D^{**}} \simeq m_B-m_D  \simeq
m_b-m_c$ and the functions $|\vec{q}{\,}|_0(t)$ and $|\vec{q}{\,}|_1(t)$, as well as $|\vec{q}{\,}|_{free}(t)$, become
identical, and the functions $L(q^0,\vec{q}~^2)$ become also identical for all states. Then, the first
contribution equates the free quark decay rate, while the second one : 
\bea \delta \Gamma_0 \simeq
\int_{|\vec{q}{\,}|_0(t=(m_B-m_D)^2)}^{|\vec{q}{\,}|_0(t=0)}~d|\vec{q}{\,}|~|\vec{q}{\,}|^2~L_{n=0}(|\vec{q}{\,}|)~(-\rho^2~{|\vec{q}{\,}|^2 \over m_b^2})
\label{isgur0}, 
\eea 
is exactly  cancelled by the excited state contribution : 
\bea
\Gamma_1 \simeq
\int_{|\vec{q}{\,}|_1(t=(m_B-m_{D^{**}})^2)}^{|\vec{q}{\,}|_1(t=0)}~d|\vec{q}{\,}|~|\vec{q}{\,}|^2~L_{n=1}(|\vec{q}{\,}|)~\tau^2~{|\vec{q}{\,}|^2 \over
m_b^2}, 
\eea 
due to Bjorken sum rule. Whence duality. However, when quark masses are
finite, there is a small part of the  integral (\ref{isgur0}) which is uncancelled, in
spite of the Bjorken sum rule, by the corresponding excited state contribution, in particular because $t=(m_B-m_{D^{**}})^2$ now differs from $t=(m_B-m_{D})^2$. We estimate the mismatch as : 
\bea 
\delta \Gamma \simeq
\int_{|\vec{q}{\,}|_0(t=(m_B-m_{D})^2)}^{|\vec{q}{\,}|_0(t=(m_B-m_{D^{**}})^2)}~d|\vec{q}{\,}|~|\vec{q}{\,}|^2~L_{n=0}(|\vec{q}|{\,})~(-\rho^2~{|\vec{q}{\,}|^2
\over m_b^2}). 
\eea 
In  this calculation, following \cite{isgur}, we disregard all other
sources of difference, in particular the fact that the leptonic tensor functions are no
more equal, and neither are the functions $|\vec{q}{\,}|_n(t)$ for $n=0$ and $n=1$ respectively, and that also the first contribution in \ref{isgur00} no longer equates the free quark decay rate. Then, our point is that this mismatch of total
widths is very small with respect to the terms we have retained. Indeed, the integral
runs over a small part of the phase space, but in addition the integrand is much smaller
near zero recoil, where the mismatch takes place, first because of the leptonic factor $L(q^0,\vec{q}~^2)=3(q^0)^2-|\vec{q}{\,}|^2$,
second because of the factor $(-\rho^2~{|\vec{q}{\,}|^2 \over m_b^2})$. Since
$L(q^0,\vec{q}~^2)=3(q^0)^2-|\vec{q}{\,}|^2$, using $|\vec{q}{\,}|_0(t=(m_B-m_{D^{**}})^2) \simeq \sqrt{2
\Delta \over \delta m} |\vec{q}{\,}|_{0,max}$ : 
\bea \delta \Gamma \simeq {\rho^2 \over m_b^2}
 3~(\delta m)^2~{({2 \Delta \over \delta m})^{5 \over 2} \over 5} |\vec{q}{\,}|_{0,max}^5, \eea
and, relative to the free quark decay rate (i.e. contribution to $\epsilon$) : \bea
{\delta \Gamma \over \Gamma_{free}} \simeq {\rho^2 \over m_b^2} {3 \over 4}~(\delta
m)^2~({2 \Delta \over \delta m})^{5 \over 2}, \eea which is parametrically small, because
of the factor $({2 \Delta \over \delta m})^{5 \over 2}$ (since $ \Delta \ll \delta m$ in
the SV limit). In fact, in our calculation we have not retained such terms.\par

Numerically too, we find it very small, with real physical masses. It is true, as noticed
in ref. \cite{isgur}, that numerically the region of Dalitz plot which is concerned is physically not
very small, because one is far from the SV limit; with our approximative formula, we find
around $20\%$ of the free decay rate in this region of phase space, not far from the
$30\%$ estimated in ref. \cite{isgur} ; but the factors considered above nevertheless combine to yield
a very small effect for ${\delta \Gamma \over \Gamma_{free}}$, around $10^{-3} \rho^2$.
This is due to the fact that the factor $(-\rho^2~{|\vec{q}{\,}|^2 \over m_b^2})$ is very small in
this region of phase space.\par 

\section{Conclusion}  Stimulated by the worries raised by
N. Isgur, we have noticed mismatches between the sum of exclusive decays and the free quark
total decay rate, which, considered separately, could convey the impression that quark
hadron duality between total widths is violated at order $\delta m/m_b^2$, because all
these mismatches are of this order. Let us recapitulate them~:\par 

1) The upper limit in
terms of $|\vec{q}|$ (corresponding to $t=0$) of the integrals for the ground state and
the  excited states contributions do not coincide. Therefore, the contributions from the
falloff of ground state and rise of excited states do not cancel near
$|\vec{q}|_{max}$ ($t=0$).\par 

2) The upper limit in $|\vec{q}{\,}|$ of the integrals for the ground state
contribution and the free quark decay do not coincide for similar reasons.\par 

3) The
leptonic tensors of the various contributions are different, because the function $q^0(|\vec{q}{\,}|)$
depends on the transition considered.\par 

At order ${\cal O} ({\delta m \over m_b^2})$, 1)
and 2) cancel between each other, while 3) has a zero net effect, by internal cancellation
of the differences of leptonic tensors, when integrated (taking into account the difference
in upper limits of integration in $|\vec{q}{\,}|$, near maximum recoil, is once more necessary).
\par 

It must be emphasized that even in this simple model and in the SV limit, it is by no
means trivial to check duality, because the check requires  to take into account detailed
effects, such as the dependence of ground state binding energy on the heavy quark masses
through their different radii, which itself reflects the flavor independence of the quark
potential, etc...\par 

In both cases, the cancellation occurs because of {\it Voloshin sum
rule}. The con\-si\-de\-ra\-tion of it is absolutely necessary, in addition to Bjorken one, to
demonstrate duality of total widths through summation of exclusive states at subleading
order. Note that, if we have an independent mean to demonstrate duality, for example by a
rigourous demonstration of OPE to the required order, we can use the result on the sum of
exclusive states, on the reverse, to demonstrate these sum rules. \par
{\bf Acknowledgements}\par
A.L.Y., L.O., O.P. and J.-C. R. acknowledge partial support from the EEC-TMR Program, contract N. CT 98-0169.

\end{document}